\documentclass[conference]{IEEEtran}
\usepackage[dvips]{graphicx}
\graphicspath{{../eps/}{../IMG/}{./img/}}
\DeclareGraphicsExtensions{.eps, .pdf}

\usepackage{array}
\newcolumntype{L}[1]{>{\raggedright\arraybackslash}p{#1}}
\newcolumntype{R}[1]{>{\raggedleft\arraybackslash}p{#1}}
\usepackage{multirow} 

\usepackage{url}

\usepackage[utf8]{inputenc}

\usepackage[
 style=ieee
,backend=biber
,isbn=false
,url=false
,doi=false
]{biblatex}

\bibliography{bib23022017}
\usepackage{xspace}

\usepackage{xcolor}
\usepackage{amssymb}

\newboolean{showcomments}

\setboolean{showcomments}{false} % toggle to show or hide comments

\ifthenelse{\boolean{showcomments}}
  {\newcommand{\nb}[2]{\fcolorbox{gray}{yellow}{\bfseries\sffamily\scriptsize#1}{\sf\small$\blacktriangleright${\em #2}$\blacktriangleleft$}}

  }
  {\newcommand{\nb}[2]{}

  }

\newcommand\FK[1]{\nb{FK}{#1}}
\newcommand\EK[1]{\nb{EK}{#1}}
\newcommand\KS[1]{\nb{KS}{#1}}

\newcommand\Review[1]{\nb{R_REFSQ}{#1}}

\newcommand\parhead[1]{\vspace{2mm}\noindent\textbf{{#1}}\ \ }

\urldef{\mailsa}\path|{fabian.kneer,|
	\urldef{\mailsb}\path|erik.kamsties}@fh-dortmund.de|
\urldef{\mailsc}\path|schmid@sse.uni-hildesheim.de|

\usepackage{color, colortbl}
\definecolor{Gray}{gray}{0.9}

\usepackage{floatflt}
\usepackage{float}

\usepackage{enumitem}
\usepackage{fancyvrb}
\usepackage{fvextra}
\DefineVerbatimEnvironment% 
	{MVerbatim}{Verbatim} 
	{frame=single, breaklines=true}

\usepackage{tabularx}

\usepackage{times}
\usepackage{rotating}
\makeatletter
\let\MYcaption\@makecaption
\makeatother

\usepackage[font=footnotesize]{subcaption}

\makeatletter
\let\@makecaption\MYcaption
\makeatother

\newcommand{\ProcessName}{AdaptationExplore}  
\newcommand{\Reqf}{\#RF}
\newcommand{\aReqf}{\#ARF}

\begin{document}

% \title{A Novel Approach to the Elicitation of Adaptive Requirements based on Creativity Techniques: \\ A Controlled Experiment}

% \title{Elicit Adaptive Requirements using Creativity Techniques: A Controlled Experiment.}

\title{Elicitation of Adaptive Requirements Using Creativity Triggers: A Controlled Experiment}

%\title{A Controlled Experiment on using Creativity Techniques for the Elicitation of Adaptive Requirements}

% war: Elicitation of Adaptive Requirements: A Controlled Experiment comparing two Creativity Techniques}
%\author{Fabian Kneer \and Erik Kamsties \and Klaus Schmid}

%\author{Fabian Kneer\inst{1} \and Erik Kamsties\inst{1} \and Klaus Schmid\inst{2} }

%\author{
%    \IEEEauthorblockN{anonymous author one\IEEEauthorrefmark{1} \and anonymous author two\IEEEauthorrefmark{2}}
    
%    \IEEEauthorblockA{\IEEEauthorrefmark{1}University One\\ Street 1, 12345 Town, Country\\}
    
%    \IEEEauthorblockA{\IEEEauthorrefmark{2}University Two,\\Street 2, 67890 Town, Country\\}
%}

%\author{anonymous authors}

 \author{
     \IEEEauthorblockN{Fabian Kneer\IEEEauthorrefmark{1}, Erik Kamsties\IEEEauthorrefmark{1}, and Klaus Schmid\IEEEauthorrefmark{2}}
     \IEEEauthorblockA{\IEEEauthorrefmark{1}Dortmund University of Applied Sciences and Arts\\Emil-Figge-Str. 42, 44227 Dortmund, Germany\\
     \{fabian.kneer, erik.kamsties\}@fh-dortmund.de}
     \IEEEauthorblockA{\IEEEauthorrefmark{2}University of Hildesheim,\\
     Universitaetsplatz 1, 31141 Hildesheim, Germany\\
     schmid@sse.uni-hildesheim.de}
 }

%\IEEEoverridecommandlockouts
%\IEEEpubid{\makebox[\columnwidth]{Placeholder~\copyright2021 IEEE \hfill} %\hspace{\columnsep}\makebox[\columnwidth]{ }}

\maketitle

\IEEEpubidadjcol

\begin{abstract}

Adaptive systems react to changes in their environment by changing their behavior. 
Identifying these needed adaptations is very difficult, but central to requirements elicitation for adaptive systems. 
%For suitable adaptation decisions the system needs adequate information about the system and its environmental context. 
As the necessary or potential adaptations are typically not obvious to the stakeholders, the problem is how to effectively elicit  adaptation-relevant information. %regarding the dynamic environment of an adaptive system. 
One approach is to use creativity techniques to support the systematic identification and elicitation of adaptation requirements. In particular, here, we analyze a set of creativity triggers defined % in our \DB{\ProcessName{}-}approach 
for systematic exploration of  potential adaptation requirements. We compare these triggers with brainstorming as a baseline in a controlled experiment with 85 master students. 
The results indicate that the proposed triggers are suitable for the efficient elicitation of adaptive requirements and that the 15 trigger questions produce significantly more requirements fragments than solo brainstorming.
\end{abstract}

\begingroup
\let\clearpage\relax
\section{Introduction}
\label{sec:Intro}

%Creativity techniques have gained a lot of research interest in the requirements engineering community \cite{Berry2020}. Creativity techniques are used to invent new requirements and cope with the challenge of missing requirements. %In the development of adaptive system we need to handle the problem of missing requirements and uncertainties.

Systems need to be adaptive to ensure the appropriate functionality in many different situations, while the environment and the system themselves may change. %The reason for missing requirements and dynamic environment is raising complexity of today's systems. They are connected to various systems that exchange data and functionalities. These systems are not available all the time and at all locations. .. das muss besser mit der Literatur verknüpft sein
An adaptive system must monitor itself and the environment, which we refer to jointly as \textit{situation}. Changes in the situation may then trigger adaptations of the system (behavior).
%to react to changes originating in a dynamic environment. 

\Review{How is this related to uncertainty? Uncertainty was referenced as an essential point to take into account in the Introduction. However, in the end, no link to it was made. I wonder how adaptation has been taken in the context of the paper? Moreover, specifically wrt.\ the value/relevance that uncertainty was given at the beginning.}
%The challenge for the development of adaptive systems is to maximize the understanding of the environment to cope with the existing uncertainties \cite{Ramirez2012,Esfahani2013,Mahdavi-HezavehiSara2016,Weyns2018}.

The aim of our research is to improve the systematic development of adaptive systems from a requirements engineering perspective. Thus, we are working on an RE process to derive requirements for adaptive systems with a focus on the early phases. One idea is to systematically analyze different situations to uncover requirements, especially requirements, which are adaptation-relevant.

%During this exploration potential situations are analyzed to identify missing system information, environment information, or functionalities. The idea is to focus on a single situation, explore its properties and thus enhance the knowledge and understanding of it. 

One challenge for the development of adaptive systems is to understand the triggers for an adaptation, the adaptation itself, and the dynamic behavior of the system. That is, to \textit{elicit} adaptive requirements. %The development has to ensure that a developed adaptive system is able to make the right decisions at run time. 
Usually, stakeholders are neither consciously nor unconsciously aware of them. Thus classic elicitation techniques such as interviews or observations are less effective.

%\KS{remove details of phases}

Creativity techniques have gained a lot of research interest in the requirements engineering community as a means of exploration \cite{Berry2020}. Creativity techniques are used to trigger novel insights during  elicitation and identify new requirements and understand non-obvious implications of existing requirements.

As a basis for identification of adaptive requirements our approach relies on  15 trigger questions tailored to the needs of adaptive systems.  The use of trigger questions is rather common as a creativity technique (cf.~checklist techniques in \cite{GrubeSchmid08}).

While a plethora of general creativity techniques exist, a central hypothesis of our work is:
\begin{quote}
    \textbf{(H1)} Applying specialized creativity techniques in elicitation of adaptive requirements leads to better results than the use of generic elicitation techniques.
\end{quote}

While we do not claim to address the full breadth of this question, our experiment will still shed some light on this. % KS: reformulieren

%In the context of discussing its validity we will also refer to it as RQ0.
%\KS{Hypothese unserer gesamten Arbeit: spezialisierte Creativity techniques sind effektiver als generische für das spezielle Ziel. Das war die Basis für die Definition des konkreten Ansatzes. - Überprüfung dieser Hypothese RQ0}

The main goal of our paper is to  compare adaptation-tailored creativity triggers %against a generic creativity technique.   In particular we test \ProcessName{} trigger questions. \KS{prozess formulieren}
%our exploration approach based on the developed questionnaire 
with the generic creativity technique \textit{solo brainstorming} as a baseline in a controlled experiment. The result of applying a creativity or elicitation technique is a number of requirements \textit{ideas} or requirements \textit{fragments}, but usually not elaborated requirements. This is taken into account in the following research questions that we pursue:

\begin{description}
    \item[RQ1:] Is our approach relying on trigger questions more effective regarding the number of collected requirement fragments?
    \item[RQ2:] Is our approach perceived as more useful than brainstorming from a user  perspective?
    \item[RQ3:] How effective are the individual trigger questions 
    % of \ProcessName{} 
    objectively in terms of collected requirement fragments?
    \item[RQ4:] What is the individual usefulness of %the \ProcessName{}  
    trigger questions from a user perspective?
    
    %\item{} \textbf{RQ3:} What is the individual utility of the \ProcessName{}  trigger questions?  (a)  objectively in terms of collected requirement fragments (b) subjectively in terms of usefulness. 
\end{description}

The remainder of the paper is structured as follows. First we give an overview of related work in Section \ref{sec:relWork}. Then we introduce the two techniques used by the participants in Section \ref{sec:process}. In Section \ref{sec:expDesign} we discuss the design and materials of our controlled experiment. The results are presented and interpreted in Section~\ref{sec:results}. The threats to validity relating to our experiment are discussed in Section~\ref{sec:TtV}. Finally, we conclude and present some resulting research directions in Section~\ref{sec:conclusion}.
\section{Related Work}
\label{sec:relWork}

%In today's requirements engineering the importance of creativity is well known \cite{Berry2020} and many techniques have been successfully applied \cite{JaramilloFranco2016}.

%\KS{Adaptive requirements, creativity (empirical studies) und Elicitation (empirical studies) behandeln}

% EK: Fokus hier auf empirischen studien lassen, sonst wird es schnell ausufernd (welche Art von Wissen benötigen wir (known knowns, unknown knows, unknown unknowns), mit welchen Techniken lässt sich dieses Wissen beschaffen?). Das haben wir bislang nicht aufgearbeitet. 

Traditional elicitation techniques have been empirically studied for quite some time. A systematic review of empirical studies \cite{DavisTHJM06} summarizes that semi-structured interviews were found most effective in studies, while other techniques such as thinking aloud tend to be less effective. A single study covers also the elicitation of environmental information \cite{FowlkesSBCS00}, yet in the highly specialized domain of avionics. 

More recently creativity techniques came into focus of RE research as they can be successfully applied to elicit requirements. We focus on \textit{lightweight} techniques \cite{SutcliffeS13} here in the sense of requiring less effort for training and application.

Burnay et al.~\cite{Burnay2016} investigate the influence of creativity triggers on the elicitation of requirements. They identified six new creativity triggers, which guide stakeholders to discover new requirements associated with a particular quality of a product. \textit{Entertaining} triggers used to discover fun or captivating features, \textit{light} triggers to simplify the solution, \textit{adaptable} triggers to replace multiple products with one adaptable product, \textit{economical} triggers to reduce the consumption of resources, \textit{complete} triggers to make a solution more integrated and more comprehensive, and finally, \textit{durable} triggers to find features that make a solution more durable and long-lasting. 

Schmid~\cite{Schmid06} reported on the benefits of the deconstruction technique and trigger lists as creativity techniques to identify ideas for new product concepts. An interesting result of this industrial case study was that new concepts were discovered in a creativity session although the organization already thought of new concepts for quite some time. 

A number of controlled experiments to collect sound empirical evidence have been conducted. Berry, Herrmann, and Mich applied EPMCreate in several experiments. In \cite{HerrmannMB18} the authors compared two variants of the technique regarding the \textit{effectiveness} in generating requirements ideas and its \textit{feasibility}. 

Niknafs et al.~\cite{NiknafsB12} empirically investigated the role of domain knowledge in idea generation during requirements elicitation. The authors conducted a controlled experiment that investigates the impact on quantity and quality of generated ideas. The quality was distinguished into raw, relevant, and feasible ideas. %The influence of participants' industrial experience and RE experience in particular was analyzed.

El-Sharkawy et al.~\cite{ELSharkawy2011} provide an heuristic approach to support product innovation in RE. They suggest heuristics to derive creative requirements from idea maps. The results of a controlled experiment showed that six heuristics performed significantly better than random stimuli.

%This are some successful approaches that have developed or used creativity triggers to produce new and novel ideas. The approaches were used to develop ideas for a general system and concept of product. In our experiment we focus on the influence of creativity techniques, brainstorming and trigger questions on the exploration of dynamic environments of an adaptive system.
\FK{gerade ziehen}
Sutcliffe et al.~\cite{SutcliffeS13} investigate challenges for elicitation. In particular lightweight approaches to elicit/invent requirements using creativity techniques have been described and applied in case studies and controlled experiments. However their application to develop requirements for \textit{adaptive} systems and an empirical evaluation has not been conducted so far to the best of our knowledge. 

%Effective Requirements Elicitation in Product Line Application Engineering – An Experiment
%Sebastian Adam and Klaus Schmid \cite{adam2013}

\section{Techniques}

\label{sec:process}
\KS{Warum erklären wir das}
We will now discuss the techniques applied in the experiment. First, we provide some background on the elicitation of adaptive requirements and, in particular, we introduce the notion of a \textit{situation}. %Our work on the elicitation of adaptive requirements is embedded into a larger effort to build a continuous requirements engineering process for the development of adaptive systems, which is discussed first. %First, we discuss the background, that is the \ProcessName{} process. 

\subsection{Background}
\label{subsec:background}
\EK{Terminologie minimieren und ggfs. erklären warum der Begriff Kontext NICHT verwendet wird. Situation / Environment / Adaptive (Adaptive Requirements)}

We use the term \textit{adaptive requirements} to denote ``requirements that encompass the notion of variability associated to either a functionality or a system quality constraint'' as defined by Qureshi and Perini \cite{QureshiP09}. According to their definition, an adaptive requirement includes a \textit{monitoring specification} that takes into account the variability in the environment, \textit{evaluation criteria} and \textit{alternative behaviors} to be adopted at runtime by the software system.

% There is a lack of support for the early phases in the development of an adaptive system, which was recognized by % a recent SLR of the authors of this paper \cite{kneer2020environment}, but also by others as 
% Kneer et al. \cite{kneer2020environment} and Dey and Lee \cite{DEY2017}. 

De Lemos et al.~\cite{DeLemosGieseMueller+13} identified challenges for adaptive systems. One of these is the need for a development process for adaptive systems.  More precisely, there is a lack of support for the early phases in the development of an adaptive system, which was also recognized by Dey and Lee \cite{DEY2017} and Kneer et al.~\cite{kneer2020environment}.

%Most of the research in RE on adaptive systems focuses on decision making and the representation of adaptive requirements. %, especially at runtime. 
%Thus, our overall goal is to build a continuous requirements engineering process for the development of adaptive systems with a focus on the early phases. An overview of this process is shown in Figure \ref{fig:processdetail}. It consists of three phases. 

%\begin{figure}[tb]
%	\centering
%	\includegraphics[width=0.9\linewidth]{Images/Process_Detail2-Short.png}
%	\caption{Overview of \DB{\ProcessName{}} process}
%	\label{fig:processdetail}
%\end{figure}

%The objective of the \textit{Initial Phase} is to define goals, tasks, and resources of the adaptive system. The result is a goal model, which is used in the \textit{Exploration Phase} to identify different situations of the system. 

We introduce in the following the concept of a \textit{situation}. We believe this to be very useful in supporting the elicitation of adaptive requirements as it helps to deal with the complexity that arises from the real world. 
\KS{übergang/ Begründung}
Focusing on one situation at a time simplifies this complexity as it  \textit{slices} the problem domain into less complex chunks. We assume this to be beneficial to the systematic elicitation  of adaptive requirements by reducing cognitive load. 

A \textit{situation} can be regarded as a special kind of \textit{view}. Similar to the concept of views, it can be restricted to certain elements (e.g., entities from the real world) or specific kinds of information (e.g., structural vs.\ behavioral view).
In extension to the concept of a view, a situation further restricts the focus to a specific moment in time (or small time interval)\KS{context?} and, thus, implies a certain state of the system and its environment. This restricts the behavior and highlights certain entities like environmental entities or stakeholders. 
  
%The term \textit{situation} is commonly understood as something like ``the set of conditions that exist at a particular time in a particular place''\footnote{\url{https://www.macmillandictionary.com}}. 
%The term has also a distinct meaning in philosophy and psychology. In short, it has a \textit{perspective}, it \textit{frames the real world}, and it has an often implicit \textit{subject}. \KS{unklar}%In the remainder of this subsection, we define the term in the context of adaptive systems and compare it to related terms such as \textit{scenario}. 
%Similar to a \textit{view}, which is a well researched concept, a situation can be incomplete. 
%In contrast,\KS{?} a situation has connotations  of environment, location, and period of time.%\KS{ich finde das sehr unklar ... jetzt etwas klarer? ... so ganz klar habe ich es auch noch nicht im Kopf}

More precisely, we characterize a \textit{situation} by the structure, function, and behavior of a system and the relevant part of the environment at a particular location over a time period of particular interest. For instance, lets assume a smart street light that adapts to weather conditions, time, and frequency of pedestrians and vehicles. One situation of the smart street light could be an ambulance warning: The rescue center sends a warning to all street lights, which are on the path of an ambulance. Cars and pedestrians along the route will be notified by switching the color of the smart street light to red.
Involved environmental entities are cars, pedestrians, ambulance, and rescue centers. Involved system entities of the street light are movement sensor, distance sensor, light (color, intensity). The situation would persist as long as the  ambulance is approaching or close by. %Constraints on the situation are imposed by road safety regulations.  % Qualities: visibility of the light 

We use the notion of a situation as a basis for a creative process for unfolding adaptation scenarios. It is analyzed with the help of trigger questions (TQ), which guide the developer to spot missing environmental information, to identify new situations, and potential adaptations.

%Of course a situation can be modeled using UML or similar techniques.
%A structural model contains all known environmental entities and system components that are involved in a situation. A behavioral model covers the system's behavior in that particular situation. The key idea is to focus on a situation in order to analyze a behavior in detail and to enhance the understanding of the potential changes to the situation that may require adaptivity. 

%In this section we illustrate the two techniques that are used during the experiment. 
%Both processes were illustrated and taught before the experiment. -.. Fokus! Hier werden die Prozesse erklärt, der Experimentablauf wird später erklärt! 

We employ textual situation descriptions (as the ambulance warning above) and UML models as a basis to apply the creativity techniques illustrated in the following.

\subsection{Trigger Questions}

%\EK{Aus dem Prozess heraus die Motivation für die Technik ableiten}

%As a full-fledged RE process is not amenable to a controlled experiment, we picked as the core novel part of \ProcessName{}, which is the analysis of situations using a tailored creativity technique. 
%As stated in Section~\ref{sec:Intro} this leads to a fundamental hypothesis, which can be seen as 
%\begin{quote}
%\textit{RQ0}: Is a creativity technique, \textit{tailored to adaptive systems}, more effective than a generic creativity technique?
%\end{quote}
%An answer to this question is relevant to the whole RE research community, not only for the evaluation of \ProcessName{}.

%\EK{Referenz?}
%Rudyard Kipling’s 1902 “Just So Stories.” I keep six honest serving-men (They taught me all I knew); Their names are What and Why and When And How and Where and Who.

Trigger questions are well-known as a lightweight creativity technique and we used the 5W1H (Kipling) technique to systematically derive questions related to the 6 question words (what, where, when, why, who, and how).
The result is a set of 15 trigger questions (see Table \ref{tab:triggerQuestion}), which should trigger new ideas in the context of an adaptive system and its dynamic environment. The goal of the question set is to identify missing information regarding environment entities, resources, new functionalities, alternative realization of functionalities, and triggers for adaptations (i.e., the ingredients of an adaptive requirement). % unklar, was soll das sein:, and any missing information related to the development. 

\begin{table}[tb]
    \centering
    \caption{\ProcessName{} Trigger Questions}
    \begin{tabular}{p{8cm}}
        \hline
        High Level Objects\\
        \hline
         (1) What should the model element be related to?\\
         (2) Does the model element have missing restrictions?\\
         (3) What information could the model element provide for a feature and how could the model element be used by a feature?\\
         (4) How should the model element be accessed? \\
         \hline
         Activity (Functions):\\
         \hline
         (5) On what should the success of the functions depend on?\\
         (6) What should the function be related to?\\
         (7) Does the function have missing restrictions? \\
         (8) Could there be any obstructions or conflicts due to other functions? \\
         (9) What information should the function provide and for what could the information be used? \\
         (10) How could the function be accessed?    \\
         (11) Could the function fail?   \\
         (12) Could there be any reasons not to perform the function? \\
         \hline
         Environmental Quantities (Variables):\\
         \hline
         (13) What influences the variable? \\
         (14) Where could the value(s) of the variable come from? \\
         (15) What is the variable used for?\\
    \end{tabular}
    \label{tab:triggerQuestion}
\end{table}

The trigger questions are designed to be applied to UML models, that is they are applied to structural models (e.g., class diagrams) and behavioral models (e.g., sequence diagrams).
The first four questions are applied to classes, entities, and similar high level objects of a situation. The questions 4 to 12 are related to functions and the effects on the system and its behavior, needed resources, alternative realizations. The last three questions support the analysis of environmental quantities in a situation. They aim at identifying influences, resources, or calculations relevant to an environmental quantity. 

The answers to these questions are \textit{requirements fragments}, which could be, e.g., an idea, an incomplete requirement, a list of functions, or a quality aspect.
%The answers to these questions are fragmentary requirements (we call them \textit{requirements fragments} in the remainder of this paper), which could be, e.g., an idea, an incomplete requirement, a list of functions, or a quality aspect.

%TODO: Erläuterung wie mit TQs adaptive anforderungen gefunden werden.
A few examples of ideas (requirements fragments) related to the trigger questions in the ambulance warning situation are (Qx refers to the corresponding question in Table~\ref{tab:triggerQuestion}): 
%We focus on the function itself so the lamp needs to change its color to red and increase the intensity to 100\%. Q2: A missing restriction could be the visibility at day time or reflection of buildings or plans that make the color appear in a different color then red. 
Q8: In case the smart street light operates in an energy-efficient mode, an ambulance warning is activated only if there is a strong need due to intense traffic. % Q11: If the connection failed the surrounding cars and pedestrians should be notified (e.g. flashing lights like failed traffic lights).
Q12: High intensity of the light and frequent changes of the intensity should be avoided if species are nearby at particular times (e.g., during mating season). 
Q14: The lighting conditions could be measured with a color sensor to ensure red is really perceived as red under current lighting conditions.

%Red color - R1 Während des Tages sollten unter Umständen andere Farben verwendet werden, da Tagsüber (wenn die Sonne zusätzlich scheint) Farben nicht gut erkennt werden. Q2

%Red Color R1 Environmental conditions like visibility. The function should be eanbled only in case of poor visibility in order to save electricity. Q8

%R2 Red Color Die aktuelle Farbe der Lampe kann über einen zusätzlichen Sensor geprüft werden q14

%Road safety - R5 The function could fail because of connectivity failures. Could fail when multiple users try to access a parking space. The parkingService Provided database is unavailable. The failure could be caused by muliple reasons such as weather, failures in servers etc. On such situations a warning or notification should be communicated to the users. Q11

%Color Red intensity - R6 Feature: Connection with the "Umweltbundesamt", If a species has certain critical times (e. g.  during the mating season) and is hindered by the light, special colours or intensities during that time could help to mitigate the damage on this species Q12

\subsection{Solo Brainstorming}

We selected solo brainstorming as a generic creativity technique, because it is well-known and easy to use, and again lightweight. Solo brainstorming is a variation of brainstorming, in which only one person uses the technique \cite{Aurum1998}. Each participant  tries to elicit missing information about the system and its environment. The task was to analyze UML models (structural and behavioral models) to answer the questions ``What additional features can be added?'' and ``What environmental entities are needed for the new features?''. During the analysis the participants were told to keep the following  principles (in \textit{italics}) in mind.

\begin{LaTeXdescription}
\item[\textit{Go for quantity}] 
This means, one should focus on divergent production under the assumption that a maximum quantity breeds quality.
\item[\textit{Withhold criticism}] 
In brainstorming, criticism of generated ideas should be put ``on hold''. Instead, a participant should focus on extending or adding to their own ideas. 
\item[\textit{Welcome wild ideas}] 
To get a good, long list of suggestions, wild ideas are encouraged. Going against widely accepted assumptions is actually encouraged.
\item[\textit{Combine and improve ideas}] 
When searching for new ideas one can combine and improve ideas in any way imaginable as this may create more and better ideas.
\end{LaTeXdescription}
%`\textit{`Go for quantity}'': This means of enhancing divergent production, aiming at facilitation of problem solution through the maxim quantity breeds quality. The assumption is that the greater the number of ideas generated the bigger the chance of producing a radical and effective solution. 
%``\textit{Withhold criticism}'': In brainstorming, criticism of ideas generated should be put "on hold". Instead, a participant should focus on extending or adding to her own ideas. By suspending judgment, a participant will feel free to generate unusual ideas. ``\textit{Welcome wild ideas}'': To get a good, long list of suggestions, wild ideas are encouraged. They can be generated by looking from new perspectives and suspending assumptions. These new ways of thinking might give better solutions. ``\textit{Combine and improve ideas}'': As suggested by the slogan ``1+1=3'' it is believed to stimulate ideas by a process of association.

The result of the application of solo brainstorming is again a list of textual requirement fragments. % similar to the \ProcessName{} trigger questions.
\section{Experiment Design}
\label{sec:expDesign}
%The goal of the controlled experiment is to compare the two processes in their ability explore the environment that is to spot adaptation points. \EK{...  Ziel MESSBAR definieren} % designed to analyze a set of questions and the usefulness to explore the environment during the development of adaptive systems. For this purpose we gathering requirements and adaptive requirements. The participants work on two different examples with the previous introduced processes.

The goal of this experiment is to understand the effect of creativity techniques during the elicitation of adaptive requirements. 
%\KS{ich habe den Teil hier mit den Variablen zusammengezogen: das war fast zwei mal direkt identisch.}
%Two techniques, (solo) brainstorming (BS) and \DB{\ProcessName{}} trigger question (TQ), were used for  elicitation and compared to each other based on the identified artefacts. The techniques were tested on two case studies, a cyber-physical system (CPS): a public street light scenario (PSL) \cite{KneerK16} and the Feed me, Feed me app (FF) \cite{Bennaceur16}. In the remainder of this section we discuss the experiment design.
%\label{subsec:Variables}

\parhead{Variables.} For answering the research questions we identified three independent variables: \textit{creativity technique}, \textit{case study}, and \textit{order}. 
%\label{subsec:Variables}
We use the two creativity techniques \textit{\ProcessName{} trigger questions (TQ)} and  \textit{solo brainstorming (BS)}. 
%\FK{Done - App vs. CPS und dann passt die ARgumentation}
As case studies we use two systems, a cyber-physical system (CPS) that implements a \textit{Public Street Light (PSL)} and an app called \textit{Feed me, Feed me (FF)} to ensure that the results are not particular to a certain domain or system type. The third variable is the order in which the participants perform both techniques to ensure that there is no undetected influence of a learning effect between the application of the techniques.

The dependent variables are the \textit{number of requirement fragments} gathered during elicitation and the utility of the techniques from a user perspective.

%The experiment validates the exploration of the environment by using different techniques but also the quality of the developed questions. This means, how is the technique or in detail the question accepted by the participants and how effective is the technique regarding the exploration of the environment. For this purpose we defined two main hypotheses:

%In this section we want to introduce the analysis procedure for the research questions. For RQ1 we use hypothesis testing and for RQ2 and RQ3 descriptive statistics. 

%\label{subsec:Hypothesis}
\parhead{Definition of Hypotheses:} 
Based on the variables we defined the following hypotheses to answer the research questions.

To answer \textit{RQ1} we formulate the following hypotheses: 
\begin{itemize}
    \item {} \textbf{H\textsubscript{1,1}}: ``\textit{\ProcessName{} Trigger questions (TQ) produce more requirements fragments than solo brainstorming (BS).} 
    \item{} \textbf{H\textsubscript{1,0}}: ``\textit{\ProcessName{}Trigger questions (TQ) do not produce more requirements fragments than solo brainstorming (BS).}'' %There is no difference in the number of produced requirement fragments during the elicitation regarding the used technique.}''
\end{itemize}

For \textit{RQ2} we formulate the following hypotheses:
\begin{itemize}
    \item {} \textbf{H\textsubscript{2,1}}: ``\textit{\ProcessName{} Trigger questions (TQ) have a higher subjective usefulness than brainstorming (BS).} 
    \item{} \textbf{H\textsubscript{2,0}}: ``\textit{\ProcessName{} Trigger questions (TQ) do not have a higher subjective usefulness than brainstorming (BS).}'' % There is no difference in the usefulness of the two techniques from a user perspective.}''
\end{itemize}

We employed a 2x2x2 fractional factorial design. The three factors are: the technique (TQ, BS), the case study (FF, PSL), and the order of the techniques (BS-TQ, TQ-BS). A fractional design was used to build groups with a sufficient number of participants. We excluded groups that would force participants to repeat a technique or a case study to eliminate the risk of a learning effect. The participants were randomly assigned to the four remaining groups. Each group performed both techniques on both case studies. The groups applied the techniques in different order. This design allows us to study the effect of the technique, the case study, and the order in which the techniques were applied. The design is shown in Table \ref{tab:group_overview}.

\begin{table*}[tb]
\caption{One-half fractional factorial design: \ProcessName{} Trigger Question (TQ), solo Brainstorming (BS), Public Street Light (PSL), Feed me, Feed me (FF)}
    \scriptsize
    \centering
    \begin{tabular}{| c | c | c | c | c | c |}
         \hline
         Group & 1st Technique & 2nd Technique & 1st Case Study & 2nd Case Study & Number of Participants \\
         \hline
         1 & TQ &  BS & PSL & FF & 22 \\
         2 & TQ  & BS & FF & PSL & 21 \\
         3 & BS  & TQ & PSL & FF & 21 \\
         4 & BS  & TQ & FF & PSL & 21 \\
         \hline
    \end{tabular}
    \label{tab:group_overview}
\end{table*}

%\label{subsec:Metrics}
\parhead{Definition of Metrics.} To measure the two dependent variables we define the following metrics. 
\begin{itemize}
    \item{\textit{M1} number of fragments}
    \begin{itemize}
        \item[(a)] \#requirement fragments (\Reqf{})
        \item[(b)] \#adaptive requirement fragments (\aReqf{}) 
    \end{itemize}
    \item{\textit{M2}} subjective usefulness
\end{itemize}

%The metrics M1 are used to measure the number of fragments, related to RQ1. The number of fragments is the sum of \Reqf{} and \aReqf{}. Using M2, we measure the subjective usefulness of both techniques for RQ2. For RQ3 to measure the utility of individual question of the \DB{\ProcessName{}} trigger questions, we compare for (a) the gathered number of fragments (M1) for every individual question and for (b) the subjective usefulness (M2) for all trigger questions.

\noindent\begin{table*}[tb]
    \caption{Examples of Requirements Fragments}
\centering
\begin{tabularx}{\textwidth}{X|X} 
    \multicolumn{1}{c}{Requirement Fragments} & \multicolumn{1}{c}{Adaptive Requirement Fragments} \\
    \hline
    % R1
    (1) $[$Solar panels shall be attached to the street light...$]$ \textbf{Using Solar Energy} &
    % AR1
    (4) $[$The lighting of the street shall depend on...$]$ \textbf{weather changes}\\
    % R2
    (2) $[$The street light shall react on ...$]$ \textbf{Car enters highlighted parking spot.} &    
    % AR2
    (5) $[$The street light shall highlight the way to a free parking space...$]$ \textbf{depending on light bulb working; alternative: displaying in app} $[$..display the parking space in the connected app$]$\\
    
    % R3
    (3) $[$ The street lights shall be connected via wi-fi...$]$ \textbf{connections via wi-fi and bluetooth can not reach lights that are too far away (connection)} &
    % AR3
    (6) \textbf{When an ambulance is coming and at the same time someone is looking for a parking place then the street light must decide which light will be used and so the other function can obstructed.}
    \end{tabularx}
    \label{tab:artufactexamples}
\end{table*}

% We count every part that is relevant for a requirement as a requirements fragment. 
\KS{Einfach Scheama darstellen}
\textit{Number of requirements fragments (M1 (a))}: We accept functional requirements, quality requirements and constraints (according to the IREB definitions). The requirements template from Rupp et al./IREB \cite{RuppSH09, Pohl2011} serves as reference for functional requirements. Beside the legal obligation (like ``The system \textit{shall}''), the requirements template consists of a \textit{process verb} which characterizes the \textit{system activity}, an \textit{object} (plus additional details), and a logical or temporal \textit{condition}. 
Each textual statement resulting from trigger questions / brainstorming that covers parts of this classification is counted as a requirement fragment. Table \ref{tab:artufactexamples}  provides examples of real fragments (highlighted) for the public street light. For illustrative purposes, we enriched these fragments with some domain knowledge (in braces). Example 1 is a minimal requirements fragment which includes an activity and an object. Example 2 shows a condition, Example 3 a constraint. %Beside the template we also considered additional information for a functionality or a problem description that is relevant for a requirement as requirements fragment, shown in Example 3.

%  activity (a system performs the process autonomously, a system is a service for an user, or a system is depending on a third party)

\textit{Number of adaptive requirements fragments (M1(b))}: we use the definition of an adaptive requirement from Qureshi and Perini introduced in Section \ref{subsec:background} as a reference for counting. Example 4 shows a condition or event that triggers an adaptation (monitoring specification). Example 5 describes a base functionality with an alternative realization. Example 6 includes a variability in the operational context that would require an adaptation. If a textual statement covers parts of the reference definition, we count it as adaptive requirements fragment. 

%For illustrating the measurement we show some examples of actual fragments from our study in Table \ref{tab:artufactexamples}. The fragments are from the street light case study. The parts in square brackets are added by the authors to deliver the context for a fragment.

\textit{Subjective Usefulness (M2)} of the technique is a subjective impression of the usefulness as perceived by the participant and is measured as a value between 1 (``not useful'') and 5 (``very useful'').  

\parhead{Participants.} The empirical study was performed during the ``Requirements Engineering'' lecture at the University of Applied Sciences and Arts Dortmund, Germany 
in the winter term 2019/2020. 85 Master students from computer science, medical informatics, business information systems, and an international study program called ``Embedded Systems for Mechatronics'' participated in the empirical study. 
%\label{subsec:Material}

\parhead{Experiment Setting and Materials.} Each participant used his or her own notebook with an online form and a paper handout with the case study description during the experiment. We used two case studies from literature. \textit{Public Street Light} describes a smart street light, and was described in~\cite{KneerK16}. The main functions are \textit{light the street} using motion sensors to detect vehicles and humans and dynamically adjust the street light to save energy and reduce light pollution. Other functions include a \textit{parking space assistant} that sends information about the free parking spaces and highlights free parking spaces (blue color) and \textit{ambulance warning} that is change the light color to red, for other cars to secure the path of an ambulance.

\textit{Feed me, Feed me} was created by Bennaceur et al.~\cite{Bennaceur16}. It is an IoT-based ecosystem to support food security, ensure sufficient, safe, and nutrition food to the global population. At the personal level (focus in our case study), the app monitors the user and its environment to provide suggestions on individual activities, health, and nutrition.

A double-sided handout for the participants contained an introduction to the case, a textual description of a situation, and a structural and behavioral model of the situation represented as a class and a sequence diagram, respectively. The handout frees participants from juggling between different windows on the notebook screen.

\textit{Online Form:} We used an online form for data collection. The form contained three parts. The first part contained general questions like age and previous experiences. The second and third part contained the questions related to the techniques (the order of the techniques is depending on the group). A short textual description was given at the beginning of both technique parts. For BS the form contains a text field to enter all produced fragments and a question about the usefulness of the technique. For TQ the form contains a text-field for every question to enter the produced fragments and a question about the usefulness of every individual question and the overall technique. The three parts were separated on individual pages so the participants could focus on a specific task. 

A package with the full experimental materials, scripts, and results is available online.\footnote{https://bit.ly/2ITHDwF}%\footnote{\url{https://drive.google.com/file/d/1TY84ceLg2xuEZatXdyClVvtv0qrA_TRS/}}
%%\DB{\footnote{https://bit.ly/2ITHDwF}}\footnote{https://gofile.me/5kcNL/iqzzR3FSm}.  

%\subsection{Procedure}
\label{subsec:Design}

\parhead{Procedure.} The empirical study has started with a 20 min training session that included a motivation of the experiment, a short introduction into adaptive systems and the elicitation of requirements, followed by a presentation of the experiment materials and an explanation of techniques.

%\parhead{Grouping and Distribution of Material.} 
After the training the participants were randomly assigned to four groups. The printed documents were distributed and all students tested the access to the online form. The students filled out a pre-questionnaire about their background (e.g., skills, motivation).  

%\parhead{First and second iteration.} 
%Based on their group the students had to analyze the assigned case study with the assigned technique. 
The experiment was carried out in two iterations of 45 minutes each. After 40 minutes a notification was given to finish the writing of the last content and look at the evaluation of the technique to gather for example the subjective usefulness of the performed technique. After 45 minutes the students had to switch to the next technique and case study assigned to their group. The perceived usefulness of the \ProcessName{} trigger questions were gathered from the participants, question after question. %The usefulness of the process they had performed the specific technique. 

% Nicht so wichtiges Detail (wichtig sind alle Informationen, um das Experiment zu replizieren). Because of the location and the number of participants we had to use two different rooms for the experiment. In the first room the German students were participating and in second room the international students. Both rooms received the same training, the same material, and had the same time slots. Also the groups were randomly distributed in both rooms.

%\subsection{Analysis}

% EK: in Kapitel 5 integriert, um wiederholdungen zu vermeiden.

%The statistical analysis should shown that we can reject the null hypotheses and possibly accept the alternative hypothesis. 
%For H\textsubscript{1} the independent variables are the technique  (BS or TQ), the example (PSL or FF), and the order in which the techniques are performed. The dependent variable is the number of fragments. We used a one tailed hypothesis H\textsubscript{1,1} to prove that TQ produces significantly more fragments than BS.

\section{Results} %\TBD{Seiten 3 1/2}
\label{sec:results}

%In this section we present the results of the experiment.

% \FK{richtige stelle für filtering? }
%\parhead{Preparation of the raw data.} The first step for the analysis was filtering and enhancing the format of the raw data. Here we changed the naming of the columns to represent the containing information. e.g. changing a column name from ``What is your Age'' to 'Age'. After the naming we reformatted the answers. In the online form we mentioned the participant should use a newline for every new information, but some used numbered lists without any newline or separators like comma, bar, or semicolon. 

\parhead{Participants' background.} The pre-questionnaire revealed that the participants' age was between 21 and 35 years with an average of 26 years. The overall motivation of the participants to join the empirical study was on average 4 out of 5 possible points. 32 participants had experiences in RE from industry (avg 2.8 years) and 57 participants had programming experiences from industry (avg 3.2 years). 

\parhead{Data reduction.} The raw data contained a number of defective records which were removed, because the respective participants did not adhere to the experiment procedure in one of the following ways: (1)  if a participant had taken too much time (above 2h) or started before the experiment had started (2 cases); (2) if a participant did not finish the application of a technique, that is they reported no fragments at all (6 cases); (3) if a participant did not follow the description of the techniques (e.g. trying to perform one technique on both case studies) (4 cases). 73 records remained for further analysis.

%\parhead{t-test.} The First test that we performed on the data was the t-test to identify the significant difference between the two process results. Regarding the hypothesis H1 we performed a unpaired and two-tailed t-test \cite{Wohlin2012}. We performed the test for requirements fragments, adaptive requirements, and the sum of both. The results are shown in Tab\ref{tab:tTest}.

%\begin{table}[]
%    \caption{t-Test results for BS vs QT for M1 and M2}
%    \centering
%    \begin{tabular}{|c|c|c|c|}
%        \hline
%        Variable & t & df & p-value \\
%        \hline
%        Requirements     & -6.9901 & 140.71  & 1.013e\textsuperscript{-10} \\
%        Adaptive Req    & -9.2473 & 81.76   & 2.378e\textsuperscript{-14} \\
%        Summe           & -9.22   & 121.18  & 1.101e\textsuperscript{-15}\\
%        \hline
%    \end{tabular}
%    \label{tab:tTest}
%\end{table}

%There is a significant difference in identified requirements, adaptive requirements, and the sum of both for the techniques trigger question and brainstorming. The p-value is very low so the results are highly significant. The difference is shown in Figure  \ref{fig:resultsReqArti}(b), \ref{fig:resultsAReqArti}(b), and \ref{fig:resultsSumArti}(b). 

\subsection{Effectiveness}

The following discussion of results is structured along our research questions.

\parhead{\textbf{RQ1}: ``\textit{Is our approach relying on trigger questions more effective regarding the number of collected requirement fragments?}''} We first present the descriptive statistics. Figure~\ref{fig:sreq_T} shows the number of requirements fragments (\Reqf{}) produced by using the two techniques. Figure~\ref{fig:sareq_T} shows the number of adaptive requirements fragments (\aReqf{}) produced by technique. Trigger questions produced more fragments than brainstorming in both cases. Figure~\ref{fig:req_C} shows \Reqf{} and \aReqf{} produced by case studies. Slightly more fragments were elicited  for the FF case study.
Figure~\ref{fig:req_O} shows \Reqf{} and \aReqf{} produced by order (BS$\rightarrow$TQ or TQ$\rightarrow$BS). The figures indicate that the order of techniques did not affect the number of elicited fragments.

\begin{figure*}[tp]
  \centering
  \begin{subfigure}{0.45\textwidth}
    \centering
    \includegraphics[width=0.9\textwidth]{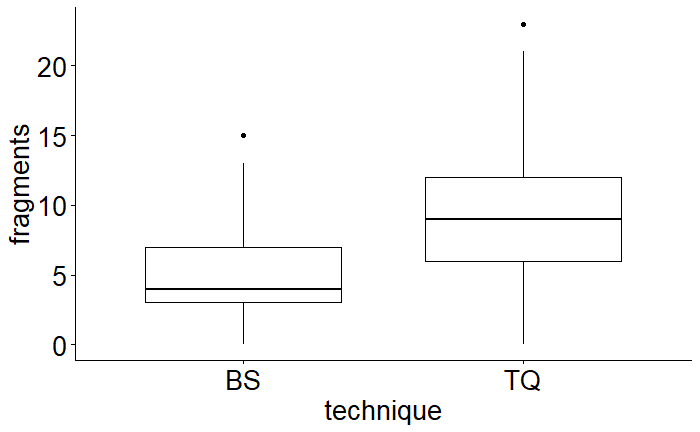}
    \subcaption{\Reqf{}}
    \label{fig:sreq_T}
  \end{subfigure}
  \hfill
  \begin{subfigure}{0.45\textwidth}
    \includegraphics[width=0.9\textwidth]{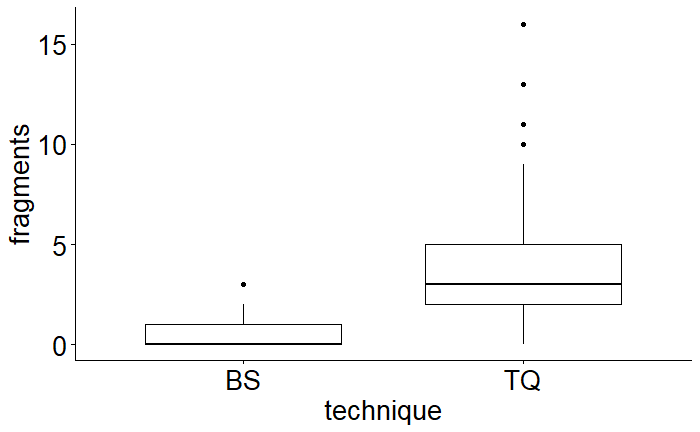}
    \subcaption{\aReqf{}}
    \label{fig:sareq_T}
  \end{subfigure}
  \caption{Fragments produced by techniques}
  \label{fig:req_T}
\end{figure*}

\begin{figure*}[tp]
  \centering
  \begin{subfigure}{0.45\textwidth}
    \centering
    \includegraphics[width=0.9\textwidth]{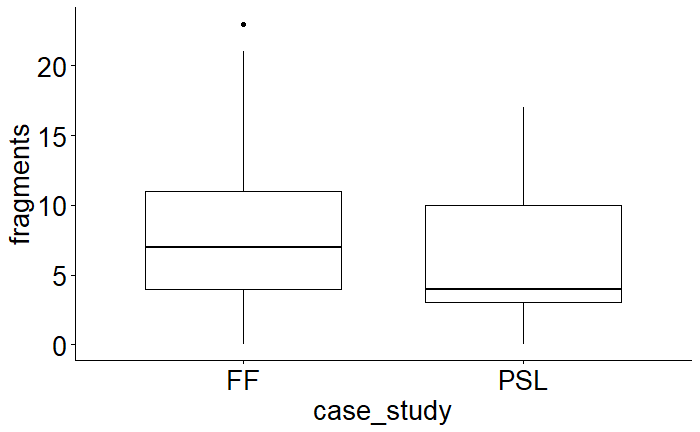}
    \subcaption{\Reqf{}}
    \label{fig:sreq_C}
  \end{subfigure}
  \hfill
  \begin{subfigure}{0.45\textwidth}
    \includegraphics[width=0.9\textwidth]{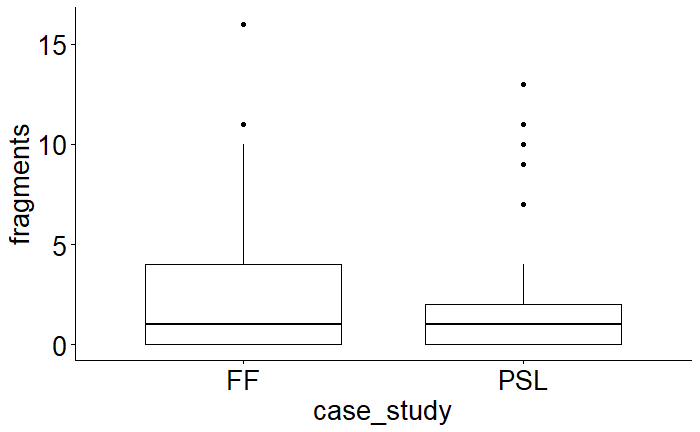}
    \subcaption{\aReqf{}}
    \label{fig:sareq_C}
  \end{subfigure}
  \caption{Fragments produced by case study}
  \label{fig:req_C}
\end{figure*}

\begin{figure*}[tp]
  \centering
  \begin{subfigure}{0.45\textwidth}
    \centering
    \includegraphics[width=0.9\textwidth]{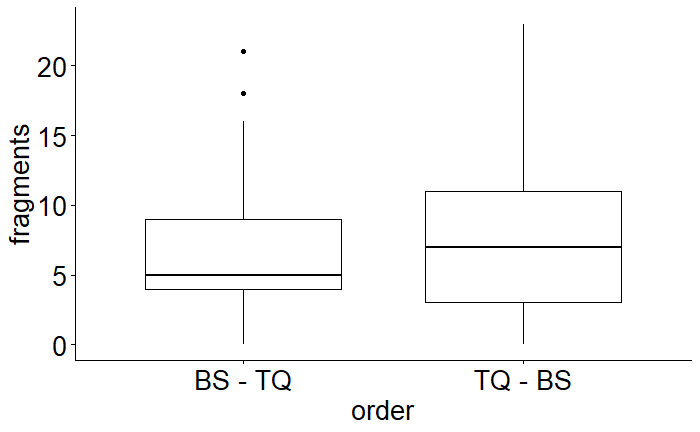}
    \subcaption{\Reqf{}}
    \label{fig:sreq_O}
  \end{subfigure}
  \hfill
  \begin{subfigure}{0.45\textwidth}
    \includegraphics[width=0.9\textwidth]{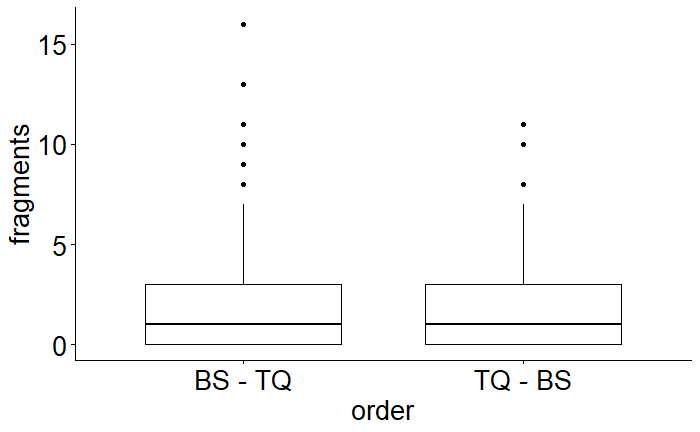}
    \subcaption{\aReqf{}}
    \label{fig:sareq_O}
  \end{subfigure}
  \caption{Fragments produced by order}
  \label{fig:req_O}
\end{figure*}

\KS{Wenn man soviel Platz für die Gefiken nimmt, sollte man auch mehr für die Beschreibung nehmen.}

The following figures compare \Reqf{} and \aReqf{} for all three factors. Figure~\ref{fig:req_TCO} shows the \Reqf{} in relation to technique, case study, and order. Figures~\ref{fig:areq_TCO} and \ref{fig:all_TCO}, show the \aReqf{} and the sum of \Reqf{} and \aReqf{}, respectively. The figures show that \ProcessName{} trigger questions produce more fragments in all cases. Moreover, \ProcessName{} trigger questions produce slightly more fragments for FF if the participants had to start with brainstorming in the first iteration.

\begin{figure}[tb] %!tb
    \centering
        \centering
        \includegraphics[width=1\linewidth]{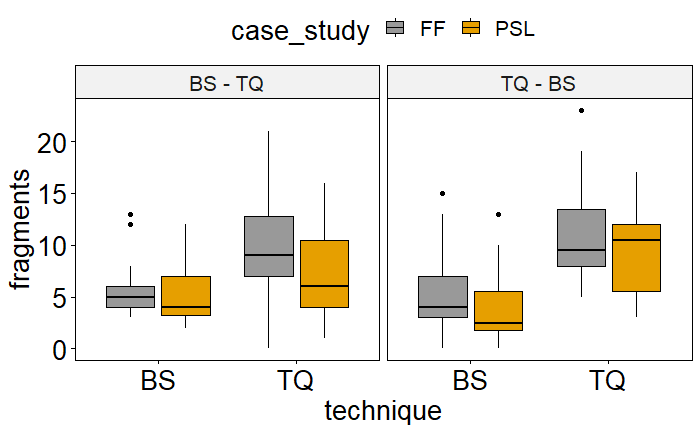}
        \caption{\Reqf{} produced by techniques related to case study and order}
        \label{fig:req_TCO}
\end{figure}
\begin{figure}[tb]
        \centering
        \includegraphics[width=1\linewidth]{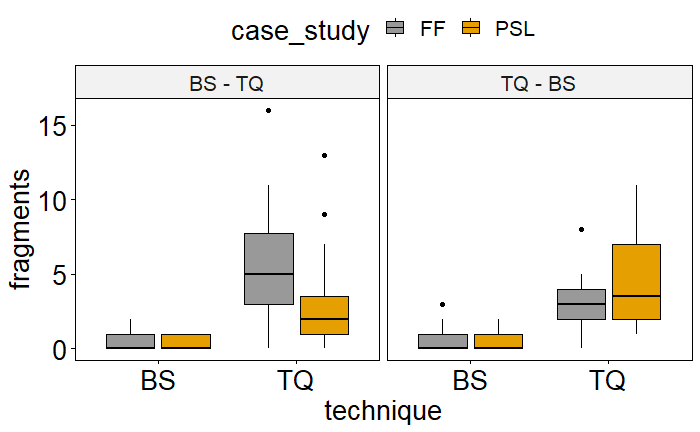}
        \caption{\aReqf{} produced by techniques related to case study and order}
        \label{fig:areq_TCO}
\end{figure}

\begin{figure}[tb] %!b
    \centering
        \includegraphics[width=1\linewidth]{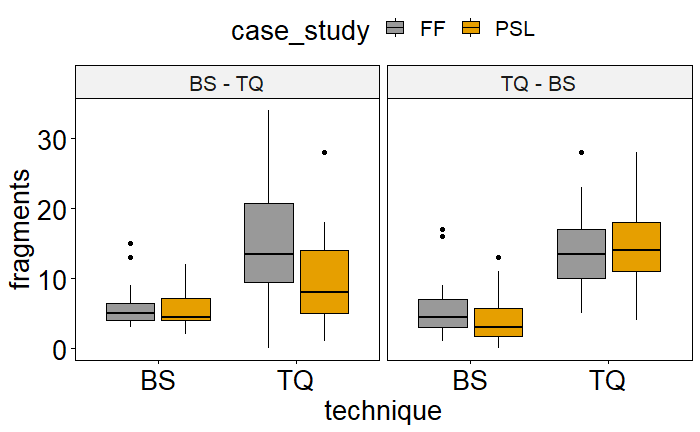}
        \caption{Sum of \Reqf{} and \aReqf{} produced by techniques related to case study and order}
        \label{fig:all_TCO}
\end{figure}

Figure~\ref{fig:reqQuestion} illustrates how many fragments were identified using a particular question. The \Reqf{} and \aReqf{} are provided for each question. The overall number of fragments varies a lot. The total number of ARFs (284) is lower than the total number of RFs (664). In most cases, the bias of a question is towards RFs, in the case of Q6, Q11, Q12, Q13 it is towards ARFs.

\begin{figure}[tb]        
        \centering
        \includegraphics[width=0.9\linewidth]{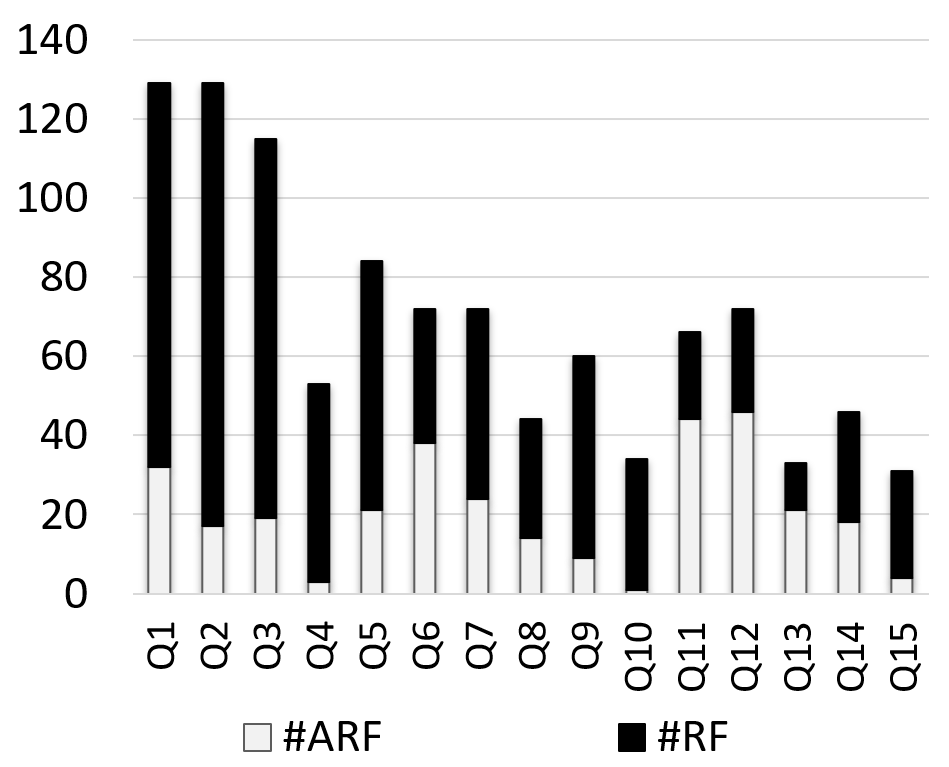}
        \caption{\Reqf{} and \aReqf{} produced by Question}
        \label{fig:reqQuestion}
\end{figure}

%The impact of each independent variable viewed in isolation is discussed below. Figure \ref{fig:resultsReqTechnique} shows the impact of \textit{technique} on \Reqf{} and \aReqf{}. For  \textit{case study} it is shown in Figure \ref{fig:resultsReqCaseStudy}, and for \textit{order} in Figure \ref{fig:resultsReqOrder}. The box plots show that the technique has an influence on the \Reqf{} and \aReqf{}, in particular \ProcessName{} trigger questions produce more fragments than solo brainstorming. The averages for case study and order are nearly the same.

%\KS{Sind eigentlich relativ uninteressant. dass is Technik + durch das Aufaddieren verschmieren Effeckte}

% \begin{figure}[htb]
%     \centering
%     \includegraphics[width=0.8\linewidth]{Images/Requirements_by_Question.png}
%     \caption{\#fragments produced by Question}
%     \label{fig:reqQuestion}
% \end{figure}

\begin{figure}[tb]
    \centering
    \includegraphics[width=0.9\linewidth]{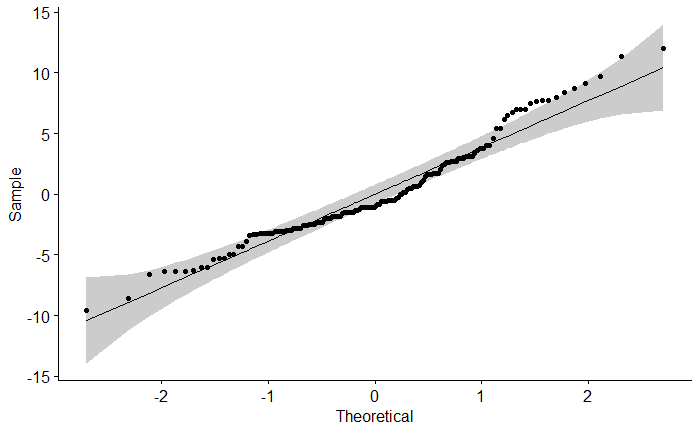}
    \caption{Quantile-quantile plot for \#RF}
    \label{fig:qqnormality}
\end{figure}

% \begin{figure}[!htb]
%     \centering
%         \centering
%         \includegraphics[width=0.7\linewidth]{Images/QQ.png}
%         \caption{Quantile-quantile plot}
%         \label{fig:qqnormality}
%\end{figure}

For the one-tailed hypothesis H\textsubscript{1} we use an ANOVA to test whether the number of fragments depends on one of the three independent variables. We use a Type III ANOVA as the group sizes are not equal. An ANOVA assumes normal distribution and homogeneity of variance. We used the Levene test to check the homogeneity of variances \cite{kassambara2019RBook}. Based on the result $p = 0.16535 > 0.05$, we can accept homogeneity. However, normal distribution for \Reqf{} is not given according to a Shapiro-Wilk test. The Q-Q plot \cite{kassambara2019RBook} provided in Figure \ref{fig:qqnormality} shows that \#RF is close to normal distribution (shaded area). As the ANOVA is reported as relatively robust against violation of normality, we proceed and provide the results of the ANOVA in Table \ref{tab:anova}.

% Based on the QQ plot we accept normal distribution as sufficiently fulfilled as all points are near to the optimal value and the ANOVA is relatively robust against violation of normality. 
%We accept the homogeneity of variances as the plot shows that there is no evident relationship between the residuals and fitted values. \FK{Nachvollziehbar erklären}

As the Levene test failed to show homogeneity for \aReqf{} and the sum of both, we applied a robust ANOVA \cite{kassambara2019RBook,MairWilcox2020} in these two cases. The results are provided in Table~\ref{tab:anova}. 

% The ANOVA was performed separately for \Reqf{}, \aReqf{} and the sum of both. 

The ANOVA shows that the technique had a significant influence on \Reqf{}, \aReqf{} and the sum of both. With respect to \aReqf{}, we can also identify a weaker significance for the case study ($p = 0.0910$), for the combination of case study and order ($p = 0.0120$), and all three variables technique, case study, and order ($p = 0.0170$). In the sum of \Reqf{} and \aReqf{} shown in the third part of the table, a significance for case study ($p = 0.0870$) and the combination of technique, case study, and order was identified ($p = 0.0790)$).

\begin{table*}[tb]
    %\scriptsize
    \caption{ANOVA Results - Significance codes:  `***' 0.001 `**' 0.01 `*' 0.05 `$\circ$' 0.1}
    \centering
    \begin{tabular}{|L{4cm}|R{2cm}|R{1cm}|R{2cm}|R{2cm}|}
        
        \hline
        \multicolumn{5}{|c|}{ANOVA Type III for \Reqf{}}\\
        \hline
        Variables & Sum Sq & DF & F value & P value\\
        \hline
        Technique                         & 154.22862	& 1 & 8.85567 & \textbf{** 0.00345}\\
        Case study                        &   2.04844	& 1 &  0.11762 & 0.73215 \\  
        Order                             &   0.39709	& 1 &  0.02280 & 0.88020 \\
        Technique:Case study              &   16.28592	& 1 &  0.93512 & 0.33523 \\
        Technique:Order                   &   11.63938  & 1 &  0.66832 & 0.41505 \\
        Case study:Order                  &   4.45675	& 1	&  0.25590 & 0.61376 \\
        Technique:Case study:Order        &   6.66140   & 1 &  0.38249 & 0.53729 \\
        Residuals                         &2403.38077   & 138 &  &\\
         \hline
         \multicolumn{5}{|c|}{Robust ANOVA for \aReqf{}}\\
         \hline
        \multicolumn{3}{| l |}{Variables}   & value & P value\\
        \hline
        \multicolumn{3}{| l |}{Technique}                   & 77.07816 & \textbf{***  0.0001} \\ 
        \multicolumn{3}{| l |}{Case Study}                  &  3.04256 &  $\circ$ 0.0910 \\
        \multicolumn{3}{| l |}{Order}                       &  0.22366 &   0.6400    \\
        \multicolumn{3}{| l |}{Technique:Case Study}        &  2.31356 &   0.1380	 \\
        \multicolumn{3}{| l |}{Technique:Order}             &  0.01363 &   0.9080    \\
        \multicolumn{3}{| l |}{Case Study:Order}            &  7.03659 & * 0.0120    \\
        \multicolumn{3}{| l |}{Technique:Case Study:Order}  &  6.34043 & * 0.0170	 \\
        \hline
        \multicolumn{5}{|c|}{Robust ANOVA for Sum of \Reqf{} and \aReqf{}}\\
        \hline
        \multicolumn{3}{| l |}{Variables}   & value & P value\\
        \hline
        \multicolumn{3}{| l |}{Technique}                      & 85.10456 & \textbf{*** 0.0001} \\ 
        \multicolumn{3}{| l |}{Case Study}                     &  3.03689 &     $\circ$ 0.0870 \\ 
        \multicolumn{3}{| l |}{Order}                          &  0.09954 &             0.7540 \\
        \multicolumn{3}{| l |}{Technique:Case Study}           &  0.67636 &             0.4150 \\
        \multicolumn{3}{| l |}{Technique:Order}                &  2.23148 &             0.1410 \\
        \multicolumn{3}{| l |}{Case Study:Order}               &  1.71474 &             0.1960 \\
        \multicolumn{3}{| l |}{Technique:Case Study:Order}     &  3.20470 &	    $\circ$ 0.0790 \\
        \hline
    \end{tabular}
    \label{tab:anova}
\end{table*}

% R docu anova The Pr(>F) gives the p value for that test.
%\KS{DISCUSSION}

As we found significant variations, we continue with a pairwise t-test to compare the means of the treatments and to accept or refute the hypothesis. We choose a more strict significance level of 0.01 (instead of the usual 0.05) as the test of several hypotheses could otherwise lead to incorrect significant values. We performed a t-test for H\textsubscript{1,1} for the technique, as the ANOVA indicated a significant difference in the variances. A pairwise t-test resulted in the following p-values: $p \le 0.0001$ for \Reqf{}, $p \le 0.0001$ for \aReqf{}, and $p \le 0.0001$ for the sum of \Reqf{} and \aReqf{}. 
Thus, we have identified a significant impact of the technique on the number of fragments in all three cases. That is, we can reject the null hypothesis and accept our hypothesis H\textsubscript{1,1}.

\parhead{Discussion.} The application of \ProcessName{} trigger questions produces significantly more adaptive requirements fragments than solo brainstorming (H\textsubscript{1}). We could also show that \ProcessName{} trigger questions produce more requirements fragments. We could not show a learning effect as the order in which trigger questions and brainstorming were applied does not have a significant impact. However, the case study had a weak significant impact. For the \textit{street light} case study, the participants reported a lower \Reqf{} than for \textit{Feed me, Feed me}. One reason could be the domain of the case study. The Feed me, Feed me case study is about a fitness gym, which might be more familiar to the participants than a public street light. Another explanation could be that Feed me, Feed me leaves more room for creativity. Additional analysis on the influences of the groups, native language, and time did not expose any significant influences on the \Reqf{} and \aReqf{}.

\subsection{Usefulness} 

\parhead{\textbf{RQ2}: ``\textit{Is our approach perceived as more useful than brainstorming from a user perspective?}''} The overall usefulness of the two techniques are shown in the last two columns of Figure~\ref{fig:helpfulall}. The \ProcessName{} trigger questions (3 out of 5) were rated as not as useful as brainstorming (4 out 5). 

% aus 4.2. relativ redundant. The results are represented as box plots and stacked bar charts for the descriptive statistics. The descriptive statistics is used to answer the research questions RQ3 (a) and (b) by illustrating the different objective results for the individual questions in form of stacked bar chart (\Reqf{} and \aReqf{}) and a box-plot that illustrated the subjective usefulness of the questions. 

\begin{figure}[tb]
	\centering
	\includegraphics[width=0.9\linewidth]{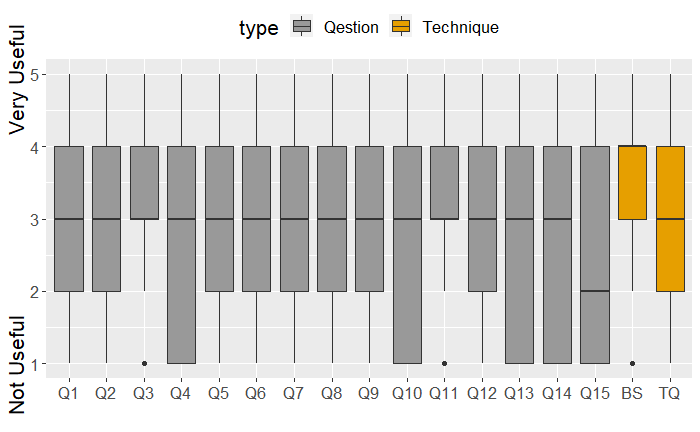}
	\caption{Usefulness of Questions and Overall Usefulness of Techniques}
	\label{fig:helpfulall}
\end{figure}
\KS{Hier kan man nicht wirklich was erkennen. Wieso sind denn durchschnittswerte..}

%\begin{figure}[htb]
%    \centering
%    \subfigure[Age]{\includegraphics[width=0.25\textwidth]{Images/age.png}}
%    \subfigure[Motivation]{\includegraphics[width=0.25\textwidth]{Images/Motivation.png}}
%    \subfigure[Motivation]{\includegraphics[width=0.45\textwidth]{Images/experience.png}}
%\caption{Results of the Participants Survey (Age and Motivation}
%\label{fig:resultsSurvey}
%\end{figure}
%\FK{Grafiken rausnehmen Text ausreichend.}

Regarding the hypothesis H2 we performed an unpaired and one-tailed t-test \cite{Wohlin2012}. We tested the usefulness of the \ProcessName{} trigger question greater than brainstorming. The result was $p = 0.9997$. This means that we need to reject H\textsubscript{2.1}. For the discussion of the results we performed a second test to analyse the effect of brainstorming on the subjective usefulness. For the second test we used a one-tailed t-test on the hypothesis BS $>$ TQ in terms of subjective usefulness, with the result $p=0.0003$. Thus, the participants even judged the usefulness of our approach as significantly lower than brainstorming.  
%\EK{Done - Neue Hypothese aufstellen, wenn die ursprüngliche gescheitert ist?! - Umschreiben nötig (Telko Freitag)}

\parhead{Discussion.} We need to reject our respective hypothesis, \ProcessName{} trigger question were regarded as less useful than solo brainstorming (H\textsubscript{2.1}). It is an interesting observation that the participants even produce more fragments using the \ProcessName{} trigger questions (objective), but rate them as less useful (subjective). 
One reason could be that the participants deem the prescriptive character of trigger questions too restrictive. Also, a single trigger question produces not as many results compared to the principle of brainstorming, which could lead to the overall impression of being less useful.

%\KS{unklar was gemeint ist}There are differences in the realization of both techniques based on an online form, the brainstorming form was more compact, which may lead to the false impression of being more useful. % One reason could be that in the online form brainstorming had a single input field, so the overall results were easier to see than the rather short results produced by the single questions.\EK{klarer formulieren}

\parhead{\textbf{RQ3}: ``\textit{How effective are the individual trigger questions  of \ProcessName{} objectively in terms of collected requirement fragments?}''}  \#RF and \#ARF are depicted in Figure~\ref{fig:reqQuestion}. The results show that the first three questions produced the most requirements fragments.

\parhead{Discussion.} RQ3: The questions were designed from being more abstract (Q1) to being quite detailed (Q15). Thus, the abstract questions seem to be more effective than the concrete ones. With respect to \#ARF, the effectiveness of the questions vary over the whole question set indicating that fatigue was not a problem. Q6, Q11, Q12, Q14 were specifically designed to spot ARFs and they actually show the best performance. 

\parhead{\textbf{RQ4}: ``\textit{What is the individual usefulness of the \ProcessName{} trigger questions from a user perspective?}''} The usefulness of each trigger question is illustrated in Figure~\ref{fig:helpfulall}. For the questions Q1 to Q14 the average usefulness was 3 out of 5. Q15 had only a usefulness of 2 out of 5.

\parhead{Discussion.} RQ4: The overall usefulness of the questions was rated \textit{medium} (3) and is in line with the overall rating of \ProcessName{} trigger questions. There are no significant differences between the questions regarding the subjective usefulness as perceived by the participants (see Figure~\ref{fig:helpfulall}). The usefulness of the last question dropped slightly when compared to the previous ones.

\section{Threats to Validity}
\label{sec:TtV}
\FK{Die Referenz auch streichen oder an den ersten Satz bauen}
Here, we will discuss some major threats to  validity \cite{Wohlin2012}. %The four main types of threats, conclusion, internal, construct, and external threats \cite{Wohlin2012} are analyzed below.

\textit{Conclusion validity} concerns the relationship between the treatment (techniques) and the outcome. We followed the assumptions of the statistical tests when analyzing the results. While we had a rather high number of participants for a Masters course, the design divided this by four leading to group sizes ranging from 16 to 20. A lack of process adherence of participants led to slightly unbalanced group sizes. 
%A group size $\ge 30$ would allow to assume normal distribution without a respective test. 

% Another threat was that some students had to leave the experiment earlier because of an external event (explosive device in city center). Due to the random separation of the participants in groups, we argue the effect is similar to both techniques and the treat can be neglected.
% \FK{Bombe auf der B1 erwähen und zu welcher art von Threat gehört es} hatten wir schon drüber gesprochen: ein ceiling effect 

\textit{Internal validity} can be affected by a number of factors. (1) The selection of subjects could be a threat. In this experiment, the students joined on a voluntary basis and were randomly assigned to groups, which should minimize selection issues. 
(2) We cannot ensure that the participants completely followed the instructions for the techniques. We tried to mitigate this threat by collecting the requirements fragments per question (TQ) / principle (BS). (3) All experimental materials are written in English. The participants had a wide variety of backgrounds. 3 native English speakers participated, most of the others (36) reported German as mother language, Kanda (10) and Tamil (5) were second and third. We tested the effect of technique and language using ANOVA, there was no significance with respect to the language. (4) An unforeseen event happened during the experiment session which caused a slight ceiling effect: some students, who came by car left the session 20-30 min before the end. This was probably caused by fear of a traffic jam as extraordinary road restrictions were announced just before the experiment started. Because of the random assignment of participants to groups, we believe this did not influence the main results very much. However, it explains to some extent the relatively low numbers of fragments found in general.

% TODO EK: der Bombenalarm hatte Einflüsse, genauer auseinander dividieren
% The ESM students have a C1 level according to the Common European Framework for Languages. The students of the other programs took a course on Technical English as part of their BSc degree, but could participate in the MSc program without any proof of English language proficiency.  
\KS{unklar}
\textit{Construct validity} concerns the relationship between theory and observation. We count the number of fragments that were spotted by the participants using the two techniques. This is common practice also in other experiments on elicitation and creativity. Yet, it is a sub-optimal measure as there is some ambiguity in how to count precisely: a participant's statement might contain one or more fragments (also called \textit{idea} in related work). We tried to reduce this threat by employing strict counting rules and by discussing ambiguous cases among all three authors. 
%, the international students need to perform a English test before they can study at the University of Applied Sciences and Arts Dortmund. % Also the Requirements Engineering lecture were hold in English. This facts should reduce the threat of the language. 
Another threat is the quality of the experiment materials. For each case study, the students received a textual description, a class diagram, and a sequence diagram. The material was introduced to the participants before the experiment to make sure the experiment material was understood by all participants. %The quality determines how easily the participants understand the material, the goal of the application, or the direction of development. 

\textit{External validity} concerns the generalisability of results. As with any study using students as subjects, the validity of the population is a concern. The goal population of this study are practitioners. The students in our study are on the Master level, with an average age of 26 years, and most of them with some industrial experience. Some of the students even have experience in requirements engineering (up to 5 years). That is, we can consider them as young professionals.% - a subset of our target population.

%During the experiment, an explosive device from a previous war was found near the main traffic artery. The local students received a notification that the road will be closed, which produce traffic jam in the whole city. The students started to work faster and try finish earlier to avoid the closing and produced traffic jam. We tested the influence of the separated groups in the two rooms and also the time spanned on the study and could find any significance.

\Review{(5) Threats to validity can mention the effect of the experiment materials: The experiment started by distributing a handout containing a piece of textual information on a situation and a set of class diagram & a sequence diagram. How far the quality of such inputs can positively and negatively create a bias to the end results of the experiments?}

\section{Conclusion}
\label{sec:conclusion}

\FK{H1 aufnehmen was können wir dazu sagen 

Applying specialized creativity techniques in elicitation of adaptive requirements lead to better results than the use of generic elicitation techniques.

}

The systematic elicitation of requirements for adaptive systems is an important part of software engineering for adaptive systems.
In this paper, we evaluated the effectiveness of a technique for elicitation of such requirements, which relies on the use of trigger questions, an approach taken from the area of creativity techniques. We crafted 15 trigger questions and evaluated them in this paper.
Our evaluation relies on a controlled experiment that compares adaptation-tailored trigger questions to solo brainstorming as a baseline. 

The results are very positive for trigger questions as the number of requirements artifacts produced (both adaptive and other) is significantly higher than with the baseline.
Beyond an analysis relying on objective metrics, we also evaluated the subjectively perceived usefulness, which gave rise to a very interesting divergence in the results as participants rated brainstorming as more useful, i.e., exactly opposite of the objective data.
We argue that this divergence is a benefit of our study as such divergences can only be detected, if both subjective and objective metrics are combined, which is only rarely the case.
We also evaluated the various trigger questions individually, seeing significant differences in their productivity.

The participants of our experiment were mostly students, but many of them spent a considerable time in industry between their BSc degree and enrollment into the MSc program. We expect that this reduces any concerns regarding the relevance of the results to industrial settings. 

%EK Overall, we believe this experiment shows that \ProcessName{} is a significant step towards the creation of better methods for requirements elicitation for adaptive systems. 

Overall, we believe this experiment shows that the use of \textit{creativity techniques} is a very important for creating better techniques for requirements elicitation for adaptive systems. Moreover, the results of our experiment are in line with the fundamental idea, expressed in by H1 in the introduction that defining \textit{specialized} creativity techniques is more effective than general ones. 

We see significant potential for future research:
\begin{itemize}
%   \item The specific triggers are a first successful step, but further research is needed for optimizing them. 
%    A basis is given in this paper as we provide first insights into the different effectiveness of individual triggers.
  \item The current list of triggers was an initial attempt, there are probably possibilities to further optimize them. 
  
%    \item The relation between specific characteristics of the case studies and the effectiveness of the triggers should be further analyzed. For the case study FF slightly more fragments were identified. We need to further analyse if the trigger questions may work better for this type of system or in this particular domain.
%    \FK{mehr details was passieren soll}
\item A better understanding of the influence of domain characteristics would be useful. For example, we observed more results for the FF study, but currently it is unclear, whether this is due to domain characteristics, interactions between questions and the domain or a statistical artifact. 

    \item Tool support could be investigated to interactively ask for a specific element, system type, or domain, based on triggers similar to the proposal by El-Sharkawy and Schmid~\cite{ELSharkawy2011}. %, following preliminary ideas from our earlier research.~\cite{ELSharkawy2011, DologLin+09}.
    %\FK{mehr details was passieren soll}
    
%    \item A very important indication for future research is the divergence between objective and subjective measurements. We need further studies to better understand the subjective ratings and construct methods that are both subjectively useful (and hence are preferred by method users) and are objectively useful.
  \item The divergence between objective and subjective measurement leads to questions like: where does this divergence come from? Can we improve the approach so that it is seen as subjectively more useful (and hence preferred by method users, while remaining objectively superior.
\end{itemize}
We regard especially the last point as very important as it provides a direct path to method adoption as we assume that independently of the objective data, only methods will be adopted by engineers that also \textit{feel right} to them, i.e., that are also subjectively productive. 
We also believe that while this divergence may look like an issue in our results, it is actually a virtue as such divergences are rarely discussed, even though, we believe they are rather common.

\endgroup

\printbibliography

\end{document}